# Insights into thermal transport property of monolayer $C_4N_3H$: a first-principles study

Yelu He[a], Dingxing Liu[a], Yingchun Ding[b,*], Jianhui Yang[c,*], Zhibin Gao [d,*]

[a]Teacher School of Education, Chongqing Three Gorges University, Wanzhou 404100, Chongqing, People's Republic of China

[b]College of Optoelectronics Technology, Chengdu University of Information Technology, Chengdu, 610225, People's Republic of China

[c]School of Physics and Electronic Engineering, Leshan Normal University, Leshan, 614004, People's Republic of China

[d]Department of Physics, National University of Singapore, Singapore 117551, Republic of Singapore

## ABSTRACT

The electronic and thermal transport properties havebeen systematically investigated in monolayer $C_4N_3H$ withfirst-principles calculations. The intrinsic thermal conductivity of monolayer $C_4N_3H$ was calculated coupling with phonons Boltzmann transport equation. For monolayer $C_4N_3H$, thethermal conductivity (κ) (175.74 and 157.90 W $m^{-1}K^{-1}$ with a and b-plane, respectively) is significantly lower than that of graphene (3500 W $m^{-1}K^{-1}$) and $C_3N$(380 W $m^{-1}K^{-1}$). Moreover, it is more than the second time higher than $C_2N$ (82.88 W $m^{-1}K^{-1}$) at 300 K. Furthermore, the group velocities, relax time, anharmonicity, as well as the contribution from different phonon branches, were thoroughly discussedin detail. A comparison of the thermal transport characters among 2D structure for monolayer $C_4N_3H$, graphene, $C_2N$ and $C_3N$ has been discussed. This work highlights the essence of phonon transport in new monolayer material.

*Corresponding authors:*

Yingchun Ding, Email: *dyccqzx@cuit.edu.cn*

Jianhui Yang, Email: *yjh20021220@foxmail.com*

Zhibin Gao, Email: *zhibin.gao@nus.edu.sg*

# 1  Introduction

Recently, two-dimensional (2D) materials such as graphene and other like-graphene have drawn considerable attention because of their excellent performance and potential applicationinenergy conversion and next-generation electronic technologies [1-3]. It is known that the thermal conductivity is the most important parameter in heat management in a wide range of technology andengineering applications [4-6]. Moreover, the property of thermal transport is a significantfactor in the application of materials [7,8]. Forexample, materials with high thermal conductivity usually are used in modern integrated circuits and electric devices to remove the accumulated heat thus ensure the stability and high-performance of the device and extend the service life, while low thermal conductivity materials are looking forwardto application in thermoelectric devices [7-13].

After the discovery of graphene, the two-dimensional (2D) materials have become one of the hot spots in the current material research community, owing to its structural stability and fascinating electronic and thermal properties with extremely high conductivity which is around 3500Wm$^{-1}$K$^{-1}$[14]. All these various applications for heat management and renewable energy conversation are closely related to the thermal transport properties. The monolayer MX$_2$ is a promising material for spintronics and valleytronics due to the remarkable high lattice thermal conductivity [11,12].

Peng et al. have investigated the thermal transition properties of 2D group-IV and group-V materials [15], and they found that the magnitude of lattice thermal conductivity fornitrogeneis comparablewith that of hydrogenated graphene (876 W m$^{-1}$K$^{-1}$), and has quitelarger lattice thermal conductivity compared to penta-graphene (255 W m$^{-1}$K$^{-1}$) [16]. The thermal conductivities of blue phosphoreneand arsenene are in the range of those of MoS$_2$ (34–155m$^{-1}$K$^{-1}$) [17-18], blue phosphorene (78 W m$^{-1}$K$^{-1}$) [19] and black phosphorene (36 W m$^{-1}$K$^{-1}$ in armchair direction, 110 W m$^{-1}$K$^{-1}$ zigzag direction) [20]. The high lattice thermal conductivity of graphene, nitrogene, phosphorene, and arsenene implies that these materials are promising candidates applied in electronic devices for efficient heat dissipation [15]. In order to find new materials with high thermal conductivity, people also studied the nitrogenated holey graphene (NHG) [21] and monolayer carbon nitride C$_3$N [22]. The calculated intrinsic lattice thermal conductivity of NHG and C$_3$N is predicted to be about 82.22W m$^{-1}$K$^{-1}$ and 380 W m$^{-1}$K$^{-1}$ respectively, at room temperature [21, 22].

A new class of 2D metal-free organic Dirac (MFOD) materials has been proposed in the past few years [23-27]. Among them, monolayer $C_4N_3H$ has been focused more and more. The monolayer $C_4N_3H$ has small high Fermi velocity, large Poisson's ratio, stiffness constant and robust Dirac cone. Therefore, the monolayer $C_4N_3H$ has an important advantage of application in high-speed flexible electronic devices[26, 27]. Understand the thermal transport of monolayer $C_4N_3H$ is very essential to the application of newly fabricated metal-organic material with monolayer $C_4N_3H$.

Motivated by the application of monolayer $C_4N_3H$, in this work, by solving the phonon Boltzmann transport equation with interatomicforce constants under first-principles calculations, weinvestigate the phonon transport property of monolayer $C_4N_3H$. Moreover, to reveal the characteristics of thermal conductivity in monolayer $C_4N_3H$, we compare thermal conductivity in monolayer $C_4N_3H$ with typical $C_2N$, $C_3N$, and graphene. As is known, the lattice vibrations or phonons are the dominant heat transport carriers, thus we focus on lattice thermal conductivity in our work.

## 1. Computational details

The calculations in this work are performed within the density functional theory implemented in the Vienna ab-initio simulation package (VASP) [28]. The Perdew-Burke-Ernzerhof (PBE) of generalized gradient approximation (GGA) with the projector augment-wave (PAW) method [29] is chosen to describe the exchange-correlation functional. The cut-off energy of the plane wave is set to 500 eV in the expansion of the electronic wave function. The Γ-centered k-mesh of 5×5×1 was employed in the first irreducible Brillouin zone. The convergence criterion for total energy and force per atom isset to $1\times10^{-6}$ eV/cell and $10^{-4}$ eV/ Å respectively. The second-order (harmonic) forces constants (IFCs) and the phonon spectrum distribution of monolayer $C_4N_3H$ are calculated with PHONOPY code at the level of harmonic approximation[30].We apply the ShengBTE code to calculate the lattice thermal conductivity with the use of the self-consistent iterative approach with the phonon Boltzmann transport equation[31]. Combining with second-order (harmonic) and third-order (anharmonic) interatomic force constants (IFCs), the phonon Boltzmann transport equation is solved with an iterative self-consistent method, which has succeededin the aspect of thermal conductivity of material [6, 10, 11-19]. We compute the second-order IFCs by the finite difference method with 4×4×1 supercell. For third-order IFCs, the 4×4×1 supercell with 2×2×1 q-meshes is used for monolayer$C_4N_3H$. Eighth nearest neighbor atoms are considered in the interactions for third-order IFCs calculation. We carry out the convergence of the thermal conductivity test about the nearest nerghbor (fourth, fifth, eighth and tenth) and q-grid (20×20×1, 30×30×1, 40×40×1, 50×50×1). Increasing the q-mesh further would change the calculated lattice thermalconductivity by less than %1.Therefore, the well-converged 50×50×1 q-meshes are used for themonolayer $C_4N_3H$. The q-grid of 50×50×1 used in thermal conductivity calculation for monolayer $C_4N_3H$.

## 3. Results and discussions

3.1. Structure and phonon spectrum

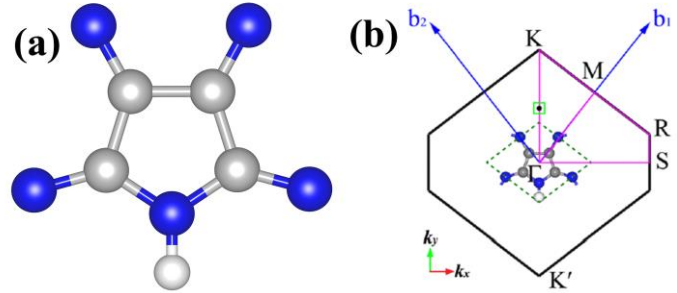

FIG. 1.(a) The optimized geometricprimitive cell of the monolayer $C_4N_3H$ on top side view; (b) the corresponding first Brillouin zone with high-symmetry k-points.

The monolayer $C_4N_3H$ has a symmetry of $C_{2v}^{14}$ and belongs to the Amm2 space group are shown in Fig 1. This novel material can be viewed as a 2D polymer network of pyrrole skeletons connected via nitrogenatoms [27, 28]. It has a rhombic primitive cell with the relaxed latticeconstants of $a_1 = a_2 = 4.769$ Å, $\gamma = 104.54$ °with PEE functional. Thesecalculated results agree with the data of Pan .et.al [26,27].

The phonon spectra (See Fig 2) and its electric band structure (See Fig 3) of themonolayer $C_4N_3H$ arecalculated based on the corresponding primitive cell in this work. In Fig. 2, we show the phonon spectrum of monolayer C4N3H along the high symmetric path Γ (0.0 0.0 0.0) - K (0.4, 0.4, 0.0) - M (0.5, 0.0, 0.0) - R (0.6, −0.4, 0.0)-S (0.5, −0.5, 0.0)-Γ (0.0 0.0 0.0) -M (0.5, 0.0, 0.0) in the first Brillouin zone. 8 atoms in unit cell, there are 24 phonon modesat each k-point.

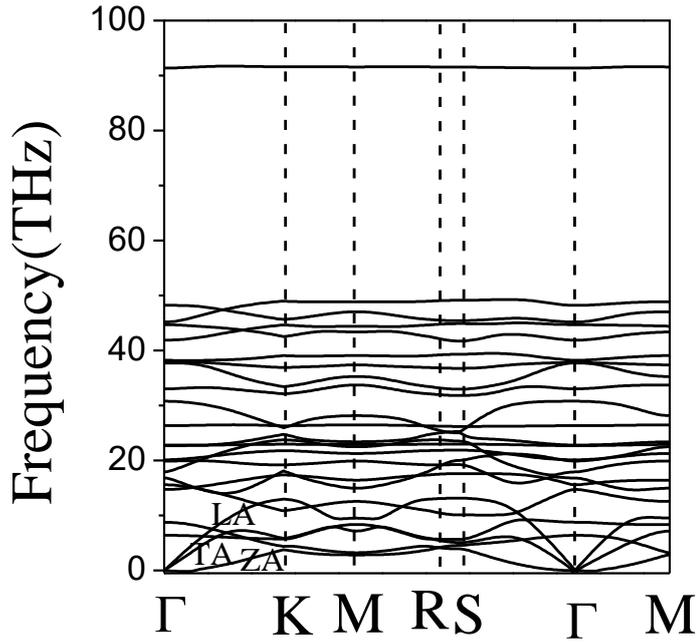

Fig. 2. The phonon spectrum of monolayer $C_4N_3H$ along with high symmetry points. First Brillouin zone with the special kpoints: Γ (0.0,0.0,0.0), K (0.4,0.4,0.0), M (0.5,0.0,0.0), R (0.6,−0.4,0.0), and S(0.5,−0.5,0.0). The longitudinal acoustic (LA), transverse acoustic (TA) and out-of-plane acoustic (ZA) branches are represented.

As shown in Fig. 2, there is no signof an imaginary phonon mode in the phonon spectrum along withthe highly symmetric points in the entire Brillouin zone. The computed phonon spectrum inagreement well with the result of Pan.et.al [26]. These results demonstratethe good dynamic stability of the $C_4N_3H$ monolayer.

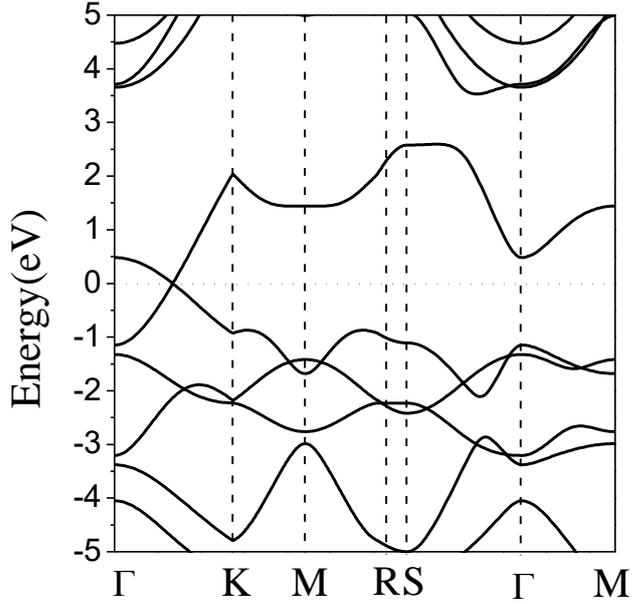

Fig. 3. Calculated band structure of the monolayer $C_4N_3H$. The Fermi level is assigned at 0 eV.

The electronic band structure of the monolayer $C_4N_3H$ has been explored and shown in Fig. 3. As shown in Fig. 3, one can see that the top of valence and the bottom of conduction bands meet along the Γ-K path at the Fermi level, which implies the monolayer $C_4N_3H$ is a semimetal. The characteristics oflinear bands and the degenerate state at this point suggest that the Dirac state exists in monolayer $C_4N_3H$. Two dimensional (2D) Dirac materials are much more desirablefor applications in nanoscale integrated circuits [23].

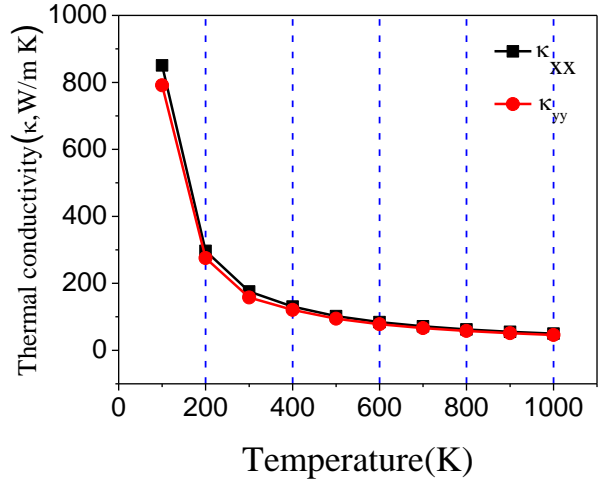

Fig. 4 The calculated lattice thermal conductivities ($\kappa_{xx}$ and $\kappa_{yy}$) of monolayer C4N3H as a function of temperature.

## 3.2. Lattice thermal conductivity

Monolayer $C_4N_3H$ belongs to the Amm2 space group with the orthogonal system [26, 27]. For orthogonal symmetry, there are two non-zero diagonal components $\kappa_{xx}$ and $\kappa_{yy}$ in lattice thermal conductivity tensor. The thermal conductivity for monolayer $C_4N_3H$ is depicted in Fig. 3 from 100 K to 1000 K. As we can see from Fig. 4, the lattice conductivities (both $\kappa_{xx}$ and $\kappa_{yy}$) decrease with the increasing of temperature, following well to the universal $1/T$ relation. Furthermore, the magnitude of lattice thermal conductivity displays slight anisotropy. The $\kappa_{xx}$ is slightly larger than $\kappa_{yy}$, especially at low temperatures. For example, the $\kappa_{xx}$ and $\kappa_{yy}$ are 175.74 and 157.90 W m$^{-1}$K$^{-1}$, respectively, at 300 K. The ratio $\kappa_{xx}/\kappa_{yy}$ equals 1.07, 1.11 and 1.08 at 200, 300 K and 1000 K, respectively. However, Pan et al. reported $C_4N_3H$ monolayer has a remarkable elastic anisotropy with ideal tensile strength [27].

In this work, we find $C_4N_3H$ monolayer has slightly different thermal lattice conductivities (the $\kappa_{xx}$ and $\kappa_{yy}$) along with a- and b-direction. In comparison with graphene, the thermal conductivity of monolayer $C_4N_3H$ is approximately one order of magnitude smaller than that ofgraphene. However, the value is relatively large compared with many other typical monolayer materials, such as blue phosphorene, $MoS_2$, etc. The thermal conductivity of $C_4N_3H$ monolayer is significant lower than that of grapheme (3716.6 [15], 3500 W m$^{-1}$K$^{-1}$[20], nitrogene (876 W m$^{-1}$K$^{-1}$) [15], $C_3N$ (380 W m$^{-1}$K$^{-1}$) [22] and 103.2 W m$^{-1}$K$^{-1}$ [32] ) are near to that of penta-graphene (255 W m$^{-1}$K$^{-1}$) [16], monolayer h-BN ( 250 W m$^{-1}$K$^{-1}$) [30] and 2D-O-silica (191.7 W m$^{-1}$K$^{-1}$) [9]. But, the lattice thermal conductivity of $C_4N_3H$ monolayer is larger than that of blue phosphorene (106.6 W m$^{-1}$K$^{-1}$)[18] monolayer $MoS_2$(100 W m$^{-1}$K$^{-1}$) [18, 33] and black phosphorene (armchair 36 (4.59) W m$^{-1}$K$^{-1}$,zigzag 110 (15.33) W m$^{-1}$K$^{-1}$) [20, 34].Therefore, to be sure, $C_4N_3H$ monolayer may performas well as or even better than other 2D-materials in various applications of thermal management, particularly in nanoelectronics.

### 3.3. Mode level analysis

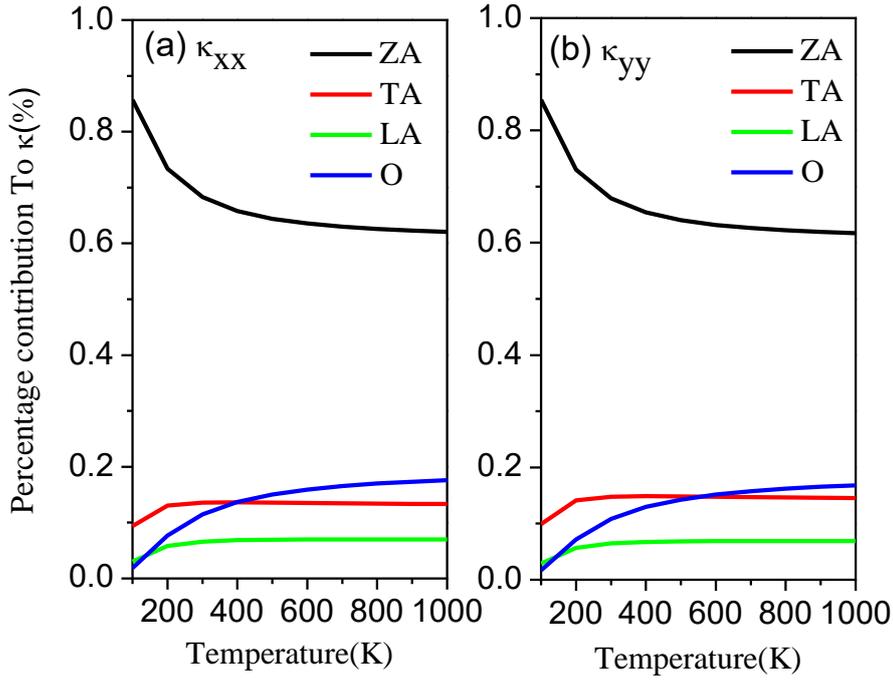

Fig. 5 Temperature dependence of thermal conductivities from different phonon branches for $C_4N_3H$ monolayer, (a) adirection; (b) bdirection.

We analyze the contribution to thermal conductivity from each phonon branch. We plot the percentages contribution of each phonon branch to thermal conductivities ($\kappa_{xx}$ and $\kappa_{yy}$) as a function of temperature from 0 K to 1000 K in Fig. 5(a) and Fig. 5(b). From Fig 5, we find that in the 0 K to 1000 K temperature range, the main contributors about not only $\kappa_{xx}$ but also $\kappa_{yy}$ of thermal conductivity are three acoustic branches.

It is noteworthy that the heat conduction is mainly conducted bythe ZA phonon branch in a- and b-direction, which is obviously in contrast to the intra-layer case. For example, the ZA branch contributes about 85% of thermal conductivityin both a- and b-direction at 100 K. With the increaseof temperature, the contribution of the ZA branch decreases gradually, reaching about 62.0 and 61.7 % to $\kappa_{xx}$ and $\kappa_{yy}$ at 1000 K, respectively. Similar behavior of the ZA branch was reported in previous heat conduction of some 2D transition metal dichalcogenides [35, 36]. We also find that the LA acoustic branch plays an unimportant role in a- and b-direction heat transport with roughly 7% from 100 K to 1000 K. For the optical phonon play the least role in a- and b-direction heat transport at low temperature. Their total optical dispersionscontributionto $\kappa_{xx}$ and $\kappa_{yy}$ is roughly 1.8 and 1.7 % at 100 K, respectively. And it is roughly 18 and 17% at 1000 K. Generally speaking, the optical phonon modes do not always contribute much to the thermal conductivity owing to their quite low group velocity and short mean free path.

We further decompose thermal conductivity into three acoustic branches (LA, TA, ZA) and all-optical phonons (O) to explore the nature of heat transport. We perform a detailed comparative analysis between $C_4N_3H$ monolayer and other existing 2D materials as shown in Table 1. In Table 1, the total thermal conductivity and the contributions from different phonon modes of monolayer $C_4N_3H$, graphene, $MoS_2$ and other 2D materials at room temperature are illustrated.

Table 1 Compared the calculated thermal conductivity ($\kappa$ in W m$^{-1}$K$^{-1}$) of monolayer C$_4$N$_3$H with other 2D materials at 300 K, the contribution from different phonon branches (LA, TA, ZA, and all-optical phonons (O) are listed together.

| | | Total thermal conductivity $\kappa$(W m$^{-1}$K$^{-1}$) | ZA(%) | TA(%) | LA(%) | Optical (%) | |
|---|---|---|---|---|---|---|---|
| C$_4$N$_3$H | $\kappa_{xx}$ | 175.74 | 68.3 | 13.6 | 6.6 | 11.5 | This work |
| | $\kappa_{yy}$ | 157.90 | 67.9 | 14.8 | 6.5 | 10.8 | This work |
| Graphene | | 3716.6 | 76.4 | 14.7 | 7.9 | 1.0 | [15] |
| | | 3500 | 80.1 | 1.1 | 6.7 | 1.1 | [22] |
| C$_3$N | | 380 | 63.2 | 2.1 | 1.2 | 33.5 | [22] |
| | | 103.2 | | | | | [32] |
| Nitrogene | | 763.4 | 66.6 | 16.1 | 16.4 | 0.9 | [15] |
| Phosphorene | | 106.6 | 31.6 | 26.0 | 38.8 | 3.6 | [15] |
| Arsenene | | 37.8 | 33.9 | 24.9 | 37.7 | 3.5 | [15] |
| MoS$_2$ | | 100 | 29.1 | 30.4 | 39.1 | 1.4 | [18] |
| | | 108 | 28 | 24 | 39 | 9 | [33] |
| Black phosphorene | armchair | 36 | 28 | 33 | 12 | 27 | [20] |
| | | 4.59 | | | | | [34] |
| | zigzag | 110 | 32 | 22 | 31 | 15 | [20] |
| | | 15.33 | | | | | [34] |
| Blue phosphorene | | 78 | 26 | 27 | 44 | 3 | [19] |
| C$_2$N | | 82.88 | 48.8 | | | 44.0 | [21] |

As listed in table 1, for monolayer C$_4$N$_3$H, the acoustic phonon modes dominate the thermal conductivity contributor, especially the ZA phonon mode, which carries 68% to the total thermal conductivity at room temperature (300 K). The thermal conductivity from ZA mode is 123.57 and 113.67 Wm$^{-1}$K$^{-1}$ in a and b direction, respectively, and that by the TA phonon modes is about 36.17 W m$^{-1}$K$^{-1}$ at room temperature (300 K). We can see that the contribution from the ZA phonon mode of $C_4N_3H$ monolayer is slightly lower that of graphene (76.4 and 81%) [15, 22]. These three acoustic phonon modes together contribute to over 89.5 and 90.2 % of the total thermal conductivity ($\kappa_{xx}$ and $\kappa_{yy}$) for a- and b-direction respectively. The optical dispersions play a less role in a- and b-direction heat transport. Their total optical dispersions contribution to $\kappa_{xx}$ and $\kappa_{yy}$ is roughly 11.5 and 10.8 %, respectively at 300 K.

Similar phenomena have been found in the previous theoretical prediction of other 2D material, where acoustic phonon modes are the dominant carriers in the thermal conductivity. But, the contribution of optical modes cannot be ignored in some other materials [19]. Such as for $C_2N$ and $C_3N$, the contribution from the ZA and OP modes are crucial, while the TA and LA modes play an unimportant role [21, 22]. Their total optical dispersions have a major contribution of about 44, 33.5 and 27 (15) % at 300 K for $C_2N$, $C_3N$ and black phosphorene, respectively [21,22].

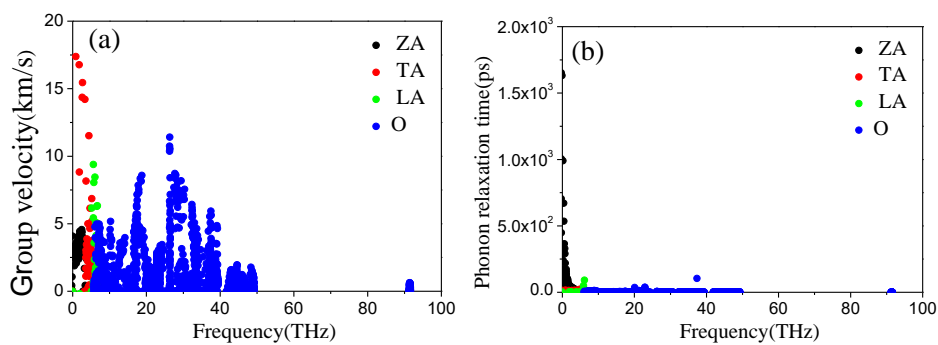

Fig. 6. Frequency-dependent the group velocity (a) and the phonon relaxation time (b) of different phonon branches at 300 K for $C_4N_3H$ monolayer.

Compared with other typical 2D materials, the group velocities of monolayer $C_4N_3H$ are considerably lower than that of graphene, $C_2N$ and $C_3N$ [15, 21, 22, 32]. It is found that the group velocities of TA and LA of $C_4N_3H$ monolayer at the long-wave length limit are about 17.5 and 10 km/s, respectively, while, the corresponding data for graphene are about 13.8 and 22.0 km/s. For ZA mode, on the contrary, we note the group velocity of graphene is significantly larger than that of the $C_4N_3H$ monolayer. We also find the significant difference existing in the contribution of different phonon branches to the total group velocities between monolayer $C_4N_3H$, $C_2N$ and $C_3N$. On the other hand, we also compare the group velocities between graphene and $C_4N_3H$ monolayer. In light of the group velocities of optical mode, the group velocities of optical mode of monolayer $C_4N_3H$ are slightly smaller than that of graphene, but the corresponding frequency distribution of monolayer $C_4N_3H$ is much wider than that of graphene.

We examine the phonon relaxation times of different phonon modes for $C_4N_3H$ monolayer as a function to phonon frequency, illustrated in Fig. 6(b). We find the relaxation times of the out-of-plane ZA modes are the biggest for monolayer $C_4N_3H$, giving the fact of perfect plane structure and geometric symmetry. The unique relaxation times of the ZA modes make ZA modes play with the dominant contributor, and provide 63.2 % to total thermal conductivity, as discussed in the previous discussion.

Furthermore, the relaxation times of $C_4N_3H$ monolayer are compared with that of graphene about the three acoustic phonon modes. The relaxation times of the three acoustic phonon modes of $C_4N_3H$ monolayer are much shorter than that of graphene, which can be attributed to the strong mixture among the acoustic and optical phonon branches in $C_4N_3H$ monolayer. Therefore we can conclude that lower group velocity and shorter relaxation time lead to lower lattice thermal conductivity of $C_4N_3H$ monolayer, compared with graphene. This result is agreement well with that of the previous of $C_3N$ [22, 32].

3.4. Phonon scattering process

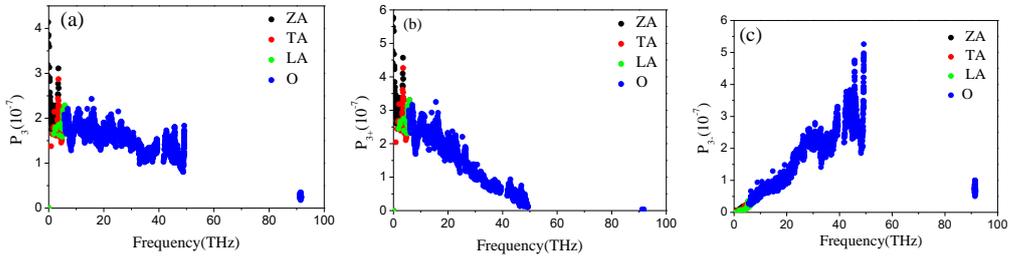

Fig. 7. Calculated phase space for three-phonon-scattering with phonon frequency at 300 K of monolayer $C_4N_3H$; total three-phonon processes ($P_3$) (a), absorption processes ($P_{3+}$) (b), andemission processes ($P_{3-}$) (c).

The phase space for the scattering of three-phonon processes and the mode Grüneisen parameter are the two crucial factors in the phonon-phonon scattering process [20,35-37]. The phase space for the scattering of three-phonon modes represents the number of channels available for phonon scattering, which is severely restricted with conservations with energy and quasi momentum conditions. The mode Grüneisen parameter represents the anharmonicity of phonon modes, which play an important role in determining the strength of each scattering channel. In order to explore the character of the phonon-phonon scattering for monolayer $C_4N_3H$, we give a discussion on the phase spaces and mode Grüneisen parameters of monolayer

C$_4$N$_3$H in more detail.

Generally, the more available scattering, the larger the scattering rate, which eventually results in the smaller thermal conductivity. The total three-phonon-scattering processes (P$_3$) consist of two independent scattering channels, the naming adsorption process (P$_{3+}$) and the emission process (P$_{3-}$). The Calculated phase space for three-phonon-scattering, including a total three-phonon (P$_3$), absorption (P$_{3+}$) andemission (P$_{3-}$) processes with phonon frequency at 300 K of monolayer C$_4$N$_3$H are given in Fig. 7.

As shown in Fig. 7(a), we note that the monolayer C$_4$N$_3$H has a large total phase space in the frequency range, especially for the acoustic (ZA and TA) and optical phonon branches. In comparison with graphene and C$_3$N, the monolayer C$_4$N$_3$H has the largest total phase space, which is consistentwith the fact that the thermal conductivity of monolayer C$_4$N$_3$H is thesmaller than graphene (3500 W m$^{-1}$K$^{-1}$) and C$_3$N (380 W m$^{-1}$K$^{-1}$) [15, 22]. From Fig. 7(a),we also find that scattering phase space of optical phonon modes is obviously larger than acoustic phonon modes in monolayer C$_4$N$_3$H. We can see from Fig. 5(b), the low-frequency optical phonon branches also contribute greatly to the adsorption process (P$_{3+}$) in monolayer C$_4$N$_3$H, mainly related to the three-phonon-scattering adsorption process of TA/LA/ZA + O→O. While, the A+A→A adsorption process is the primary three-phonon thermal transport process in graphene. We can clearly see from Fig. 7(c), the optical phonon branches play critical roles in the three-phonon emission process (P$_{3-}$) of monolayer C$_4$N$_3$H. The emission scattering phase spacefrom optical phonon branches in monolayer $C_4N_3H$ is significantly large, in which, the emission process O→LA/TA/L+O is significant.

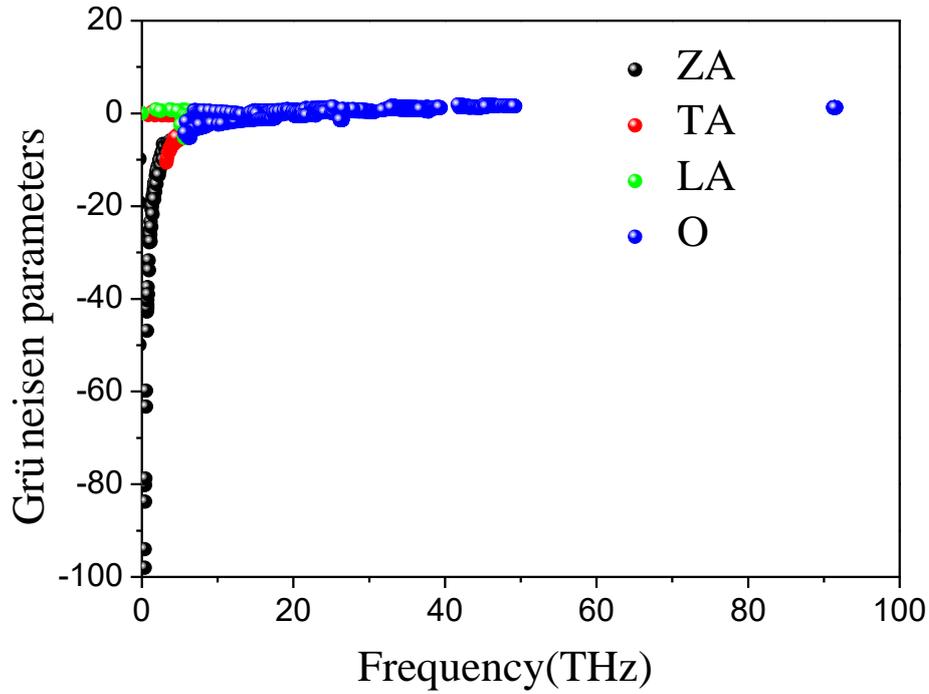

Fig. 8 Mode-Grüneisen parameter of monolayer $C_4N_3H$.

In general, the Grüneisen parameter (γ) can give the anharmonic nature of a certain structure [32-40]. We give insight into the phonon anharmonicities of monolayer $C_4N_3H$ with the Grüneisen parameter (γ). The calculated Grüneisen parameter (γ) of monolayer $C_4N_3H$ is plotted in Fig. 8. The Grüneisen parameters in Fig.8 suggests that the Grüneisen parameters of ZA branches are fully negative, while both negative and partial positive γ is found in the TA, LA and O branches for monolayer $C_4N_3H$. The large negative Grüneisen parameter value of ZA branches relates to the membrane effect in 2D materials [41-46]. However, it is worth mentioning that the scattering of ZA branches is largely suppressed in 2D materials due to the limitation of symmetryselection rule [32, 41-46]. As a result, the third-order force constant, including an odd number of out-of-plane directions disappears. In other words, neither ZA + ZA↔ZA nor ZA + LA/TA↔LA/TAscattering processcan happen in 2D materials.

We find, and what's more, the value of Grüneisen parameter ($\gamma$) in monolayer $C_4N_3H$ is a bit larger than that of graphene and $C_3N$, especially in respect ofthe $\gamma$ about three acoustic phonon branches. The larger Grüneisen parameter ($\gamma$) indicates stronger phonon anharmonicity in monolayer $C_4N_3H$, result in stronger phonon-phonon scattering. As a result, stronger phonon-phonon scattering leads to smaller phonon lifetime in monolayer $C_4N_3H$ compared to graphene and $C_3N$ [15, 22, 32 ], as shown in Fig. 4, and eventually lead to the lower lattice thermal conductivity as expected.

## 4. Conclusions

In this paper, with the use of first-principles calculations coupled with the phonon Boltzmann transport equation, we studied the phonon transport property and lattice thermal conductivity of monolayer $C_4N_3H$ sheet. Our result implies that the lattice thermal conductivity of monolayer $C_4N_3H$ is 175.74 and 157.90 W m$^{-1}$K$^{-1}$ in a- and b-directions at 300 K. We found that the heat transfer in monolayer $C_4N_3H$ is slightly anisotropic. The three-phonon process in monolayer $C_4N_3H$ was further investigated. The result shows that the absorption scattering processes between the acoustic and optical phonon (ZA/TA/LA+O→O) arevital scattering in the phonon transport for monolayer $C_4N_3H$. The calculated Grüneisen parameter (γ) of monolayer $C_4N_3H$ suggest the all the Grüneisen parameters for ZA phonon branch are negative, while both negative and positive Grüneisen parameters exist for TA, LA and O phonon branches. The phonon anharmonicity of monolayer $C_4N_3H$ is greatly higher, compared with many other typical 2D materials.

**Acknowledgments**

We would like to thank the financial supports from the Scientific Research Foundation of the Education Department of Sichuan Province (No. 2017Z031 and No. 18CZ0030), LeShan Normal University (No.LZD022, LZDP014). Z. Gao acknowledges financial support from MOE tier 1 funding of NUS Faculty of Science, Singapore (Grant No. R-144-000-402-114).